\documentclass{ws-ijgmmp}
\usepackage[cp1251]{inputenc}
\begin{document}

\markboth{Authors' Names}
{Instructions for Typing Manuscripts (Paper's Title)}

\catchline{}{}{}{}{}
%
%%%%%%%%%%%%%%%%%%%%%%%%%%%%%%%%%%%%%%%%%%%%%%%%%%%%%%%%%%%%%%%%%%%%

\title{RICCI CURVATURE AND QUANTUM GEOMETRY}

\author{MAURO CARFORA}

\address{Department of Physics, University of Pavia, via Bassi 6\\
Pavia,  27100, Italy\\
Italian National Group of Mathematical Physics (GNFM), \\
and Istituto Nazionale di Fisica Nuceare (INFN), Pavia Section\\
mauro.carfora@unipv.it}
%\email{mauro.carfora@unipv.it} 

\author{FRANCESCA FAMILIARI}

\address{Department of Physics, University of Pavia, via Bassi 6\\
Pavia,  27100, Italy\\
Istituto Nazionale di Fisica Nuceare (INFN), Pavia Section\\
francesca.familiari01@univesritadipavia.it}
%\email{francesca.familiari01@univesritadipavia.it} 

\maketitle

\begin{history}
\received{(Day Month Year)}
\revised{(Day Month Year)}
\end{history}

\begin{abstract}
We describe a few elementary aspects of the circle of ideas associated with a quantum field theory (QFT) approach to Riemannian Geometry,  a theme related to  how Riemannian structures are  generated out of  the spectrum of (random or quantum) fluctuations around a background fiducial geometry. In such a scenario,  Ricci curvature  with its subtle connections to diffusion, optimal transport, Wasserestein geometry and renormalization group, features prominently.
\end{abstract}


\section{Introduction: Ricci curvature}	

In this short review paper\footnote{This paper is an expanded version of the conference that one of us (MC) has presented on occasion of the 25th edition of the Italian theoretical physics conference series PAFT} we discuss some unconventional aspects that Ricci curvature still holds in store and which stress its basic role in a quantum field theory approach to Riemannian geometry. Even if the emphasis on the unconventional  somehow implies that we assume the reader familiar with the conventional properties of Ricci curvature, it is worthwhile to recall  some basic definitions involving our sponsor. Hence, in what follows, if not otherwise stated, $(M, g)$ is   a smooth compact $n$--dimensional manifold ($n\geq3$) without boundary whose tangent bundle $TM$ is endowed with a Riemannian metric $g$.  We respectively denote by $d\mu_g$, $\nabla$ and $\mathrm{R}m(g)$  the Riemannian measure, the Levi--Civita connection and Riemann curvature tensor  associated with $g$, where  $\mathrm{R}m(g)(X,Y)Z\,:=\,\left(\nabla _X\nabla _Y\,-\,\nabla _Y\nabla _X\,-\nabla _{[X,Y]}\right)\,Z$, for  vector fields $X,Y,Z\,\in \,C^{\infty}(M, TM)$ and where $[X,Y]$ denotes the commutator of $X,\,Y$. The  Ricci curvature  in the direction of the vector field $u\in C^{\infty}(M, TM)$ is defined by tracing the Riemann curvature according to $
\mathrm{R}ic(g)(u,u)\,:=\,trace_{\xi}\,\left(\xi\longmapsto  \mathrm{R}m(g) (\xi,u)u \right)$.  In terms of an orthonormal frame $\{e_a\}_{a=1}^n$, $g(e_a, e_b)\,=\,\delta_{ab}$, we have the Ricci (curvature) tensor expressed as $\mathrm{R}ic(g)(u,v)\,=\,\sum_{a=1}^n\,g\left(\mathrm{R}m(g) (e_a,u)v,\, e_a\right)$. In local coordinates,  $\mathrm{R}_{ik}\,=\,\mathrm{R}^h_{hik}$\,--\,(notice that the upper index on the Riemann tensor is lowered into the $4$--th position according to $\mathcal{R}{ijk\ell}\,:=\,\mathcal{R}^{h}_{\,\;ijk}\,g_{h\ell}$, so that $\mathrm{R}_{ik}\,=\,g^{h\ell}\mathrm{R}_{hik\ell}$). Similarly, by tracing the bilinear form $\mathrm{R}ic(g)$  over all possible unit directions $u$ we get the scalar curvature   $\mathrm{R}(g)\,:=\,\sum_{a=1}^n\,\mathrm{R}ic(g)(e_a,e_a)\,=\,g^{ik}\,R_{ik}$.\\
\\   The geometric nature of the Ricci tensor $\mathrm{R}ic(g)$ is made manifest by its  equivariance properties under the action of $\mathcal{D}iff(M )$, the diffeomorphisms group  of the underlying manifold $M$. In particular, if we denote by $\phi_t^*g$  the pull--back of the metric $g$ under a one-parameter group of diffeomorphism $[0,1]\ni t\longmapsto \phi_t\in \mathcal{D}iff(M )$ with $\phi_{t=0}=\mathrm{Id}_M$, then we get the equivariance relation  $Ric(\phi_t^*g)\,=\,\phi_t^*Ric(g)$, which in the limit $t\longrightarrow 0$ gives rise \cite{Kazdan} to the contracted Bianchi identity   $\nabla^i\,\mathrm{R}_{ik}\,\doteq\,\frac{1}{2}\nabla_k\,\mathrm{R}$.  The $\mathcal{D}iff(M )$--equivariance  is the rationale underlying the interpretation of  Ricci curvature (and the other curvatures) as a second order partial differential equation  acting on the metric tensor. In particular, according to a  carefully crafted analysis \cite{DeTurck1}, \cite{DeTurck2},  the Bianchi identity represents the basic obstruction for a symmetric bilinear form $A\in C^\infty(M, \otimes ^2TM^*)$ to be realized as the Ricci curvature of some Riemannian metric $\hat{g}$, \textit{i.e.}, the local obstruction  to solve the system of (weakly)-elliptic partial differential equations $\mathrm{R}ic(\hat{g})\,=\,A$,\,(weakly here refers to the fact that the Ricci tensor, thought of as a secon order partial differential operator acting on the metric, is non--degenerate only in the direction transversal to the $\mathcal{D}iff(M )$--orbit of $g$). A further and subtle set of geometrical properties of $\mathrm{R}ic(g)$ is related to   its scaling properties, a consequence of the fact that besides  diffeomorphisms, the metric $g$ is naturally acted upon also by overall rescalings according to $g\,\longmapsto\, \lambda\,g$\, $\forall  \lambda\in {R}_{>0}$
(in local coordinates this takes the form $g_{ik}\longmapsto \lambda\,g_{ik}$ and $g^{ik}\longmapsto \lambda^{-1}\,g^{ik}$). Correspondingly, we have the following induced scalings in the Riemannian volume, in the Levi-Civita connection, and in the associated curvatures:\, $Vol_{\lambda g}(\Sigma)\,=\,\lambda^{\frac{n}{2}}\,Vol_g(\Sigma)$,\, $\nabla ^{(\lambda\,g)}= \nabla^{(g)}$,\, $\mathrm{R}m(\lambda g)\,=\,\mathrm{R}m(g)$,\, $\mathrm{R}ic(\lambda g)\,=\,\mathrm{R}ic(g)$,\,and  
$\mathrm{R}(\lambda g)\,=\,\lambda^{-1}\,\mathrm{R}(g)$. Hence both the Riemann and the Ricci tensor are scale--invariant. 
 Diffeomorphisms equivariance and scaling properties of 
the Ricci tensor, with their PDE implications lurking in the background,  play a basic role  in discussing the critical points of the (volume--normalized) Einstein--Hilbert functional  $
\mathcal{S}_{E-H}[g]\doteq\,\mathrm{Vol}_g(M)^{\frac{2-n}{n}} \int_{M }\,\mathrm{R}(g)\,d\mu _{g}$, where (for $M$ compact) $\mathrm{Vol}_g(M)\,:=\,\int_M\,d\mu_g$.  These critical points are the Einstein manifolds, namely Riemannian manifolds whose Ricci tensor is such that $\mathrm{R}ic(g)\,=\,\rho_{(g)}\,g$ for some constant $\rho_{(g)}$. Notice that the Einstein constant $\rho_{(g)}$ scales non--trivially:\, Since $\mathcal{R}ic(g)$ is scale--invariant, we must have $\rho_{(\lambda g)} \longmapsto \lambda^{-1}\,\rho_{(g)}$.\\
\\  
A well--known geometrical interpretation of the Ricci curvature is provided by its expression in normal geodesic coordinates obtained by pulling back the Riemannian measure $d\mu_g$ to the tangent space $T_pM$, via the exponential map $\exp_p$ based at the point $p\in M$, (see \textit{e.g.} \cite{ABesse}).  If we denote by $d\mu_E$ the Euclidean measure on $T_pM$, then to leading order in the geodesic distance, this pull-back is provided by the  Bertrand-Puiseaux formula $\exp_p^*\left(d\mu_g\right) =\,\left(1\,-\,\frac{1}{6}\,\mathrm{R}_{ik}(p)\,u^i(q) u^k(q)\,+\,\ldots\right)\,d\mu_E$. It describes
the Ricci curvature at the generic point $p\in (M,g)$ as the distortion (with respect to the Euclidean measure)  of the Riemannian solid angle subtended by a small 
pencil of geodesics issued from $p$ in the direction $u=\exp_p^{\,-1}(q)$. From the point of view of geometric analysis a deeper insight on the nature of Ricci curvature  is provided by the expression of its components in local harmonic coordinates
$(U, \{x^i\};\,\triangle_g\,x^i\,=\,0)$.  These latter are  defined by requiring that each coordinate function defined on the coordinate neighborhood $U\,\subset\,M$,\, $x^k\,:\,U\,\longrightarrow\, \mathbb{R}$ is  harmonic. In particular, local solvability for  elliptic PDEs implies (compare with \cite{6}, Corollary 3.30) that  for any given  point $p\in (M, g)$ there always exists a neighborhood $U_p\subset M$ of $p$ such that 
$
\triangle_g\,x^j\,=\,
\frac{\partial}{\partial x^i}\,g^{ji}+g^{ij}\,
\frac{\partial}{\partial x^i}\,
\left(\ln\sqrt{\det\,g}\right)\,=\,0\;.
$
When passing from normal geodesic coordinates to harmonic coordinates we gain control on the components of the metric tensor in terms of the Ricci curvature rather than of the full Riemann tensor (cf. \cite{hebey} for a particularly clear comparison of these two coordinate systems from the point of view of geometric analysis). In particular,  as first stressed by C. Lanczos \cite{lanczos},  in harmonic coordinates  the components of the Ricci tensor take the suggestive form, (see \emph{e.g.} \cite{petersen}  Lemma 2.6 for an elegant computation),
\begin{equation}
\label{richarmoncoord}
\mathcal{R}_{ik}\;\underset{(harmonic)}{=}\;-\,\frac{1}{2}\,\triangle_{g} \,\left(g_{ik} \right)\,+\, Q_{ik}\left(g^{-1},\,\partial g \right)\;,
\end{equation}
where for each fixed pair of indexes $(i,k)$, the expression $\triangle_{g} \,\left(g_{ik} \right)$ denotes the Laplace--Beltrami operator applied componentwise to each $g_{ik}$ as it were a \emph{scalar function}, (in particular $\triangle_{g} g_{ik}$ is not the tensorial (rough) Laplacian $g^{ab}\nabla_a \nabla_b\,g_{ik} $ of the metric tensor $g$, a quantity that obviously  vanishes identically for the Levi--Civita connection).  In (\ref{richarmoncoord}),\,  $Q\left(g^{-1},\,\partial g \right)$ denotes a sum of terms quadratic in the components of $g$, $g^{-1}$  and their first derivatives, and whose explicit expression does not concern us here (compare with \cite{petersen}). Hence in harmonic coordinates the Ricci curvature acts as a quasi--linear elliptic operator on the components of the metric, an observation that can be exploited to prove \cite{DeTurckKazdan} that the metric tensor $g$, if not smooth, has maximal regularity in harmonic coordinates. It is also worthwhile to observe that the factor $1/2$ multiplying the Laplacian in (\ref{richarmoncoord}), is not as incidental as is typically assumed. This is related to the well--known and, in the words of Elton P. Hsu \cite{Hsu}, somewhat baffling fact that $\frac{1}{2}\,\triangle_{g}$ and not $\triangle_{g}$ is the generator of Brownian motion on $(\Sigma, g)$. This latter remark directly points to one of the subtlest meanings of Ricci curvature, related to its ubiquitous role in discussing diffusion over Riemannian manifolds.

\section{Ricci curvature and Riemannian manifolds with density }
Both the harmonic coordinate relation (\ref{richarmoncoord}) as well as the normal geodesic coordinates expression
$Ric_p(u,u)\propto \exp_p^*(d\mu_g)/d\mu_E $,\,$u=\exp_p^{\,-1}(q)$, are prescient signals that Ricci curvature 
may find a deeper interpretative framework in the more general setting of $n$--dimensional compact  Riemannian manifolds with density \cite{grigoryan}, \cite{gromov}. This is the set of smooth orientable manifolds $(M, g,\,d\omega)$ without boundary endowed with a  Riemannian metric $g$ and a positive Borel measure such that $d\omega\,<<\,d\mu_g$.  The absolute continuity requirement is with  respect to  the   Riemannian volume element ${d\mu_g}$,\, \emph{i.e.}, $d\omega\,=\,e^{\,-f}\,d\mu_g$, for some function $f\,\in\,C^\infty(M,  \mathbb R)$. In such a framework, the relevant  differential operators on $(M,g, d\omega)$ are the $d\omega$--weighted divergence  $\nabla^{(\omega)}_k\,v^k:=e^f\,\nabla_k\,\left(e^{-f}\,v^k\right)$, where $v^k$ are the components of a smooth vector field $v\in C^\infty(M, TM)$, and
the $d\omega$--weighted Laplacian
 $\Delta^{(\omega)}_g\,\psi:=\left(\Delta_g-\nabla{f}\cdot\,\nabla \right)\,\psi$, for $\psi \in C^\infty(M, \mathbb{R})$ \cite{grigoryan}, where $\Delta_g$ denotes the Laplace--Beltrami Laplacian on $(M,g)$. In a similar vein,  the role of the Ricci tensor is taken over by the Bakry--Emery Ricci curvature
$$Ric^{BE}(g,d\omega):=\,\mathrm{R}ic(g)\,+\,Hess_g\,f\,=\,\mathrm{R}ic(g)\,+\,\frac{1}{2}\,\mathcal{L}_{\nabla f }\,g \;,$$
where $Hess_g\,f\,=:\nabla\,\nabla\,f$ denotes the Hessian of $f$ and $\mathcal{L}_{\nabla f }g$ is the Lie derivative of the metric $g$ along the gradient vector field $\nabla\,f$. We also have a natural generalization of 
the contracted Bianchi identity $
\nabla^i\,\mathrm{R}_{ik}\,\doteq\,\frac{1}{2}\nabla_k\,\mathrm{R}$ which, for a Riemannian manifold with density $(M, g,\,d\omega=\,e^{\,-f}\,d\mu_g)$, takes the form
$$
\nabla_{(\omega)}^i\,\mathrm{R}^{BE}_{ik}\,\doteq\,\frac{1}{2}\,\nabla_k\,\mathrm{R}^{Per}\,,
$$
where
$$
\mathrm{R}^{Per}(g)\,:=\,\mathrm{R}(g)\,+\,2\,\Delta_g\,f\,-\,\left|\nabla f  \right|^2_g\,=\,\mathrm{R}(g)\,+\,2\,\Delta^{(\omega)}_g\,f\,+\,\left|\nabla f  \right|^2_g
$$
is Perelman's modified scalar curvature \cite{18}. In such a framework, the role of the Einstein-Hilbert functional is played by Perelman's $\mathcal{F}(g;\widetilde{f})$--energy \cite{18}   
$$
\mathcal{F}(g;\widetilde{f})\doteq \int_{M }\,\mathrm{R}^{Per}(g)\,d\omega\,=\,\int_M\,
\left(\mathrm{R}(g)\,+\,\left|\nabla f  \right|^2_g\right)\,e^{\,-\,f}d\mu_g
$$
and by the associated geometric functional defined on the underlying Riemannian manifold 
$(M, g)$ by
$$
\mathrm{F}[g]\doteq\,\inf_{\{f\,\in\,W^{1,2}(M),\,\int\,e^{-f}d\mu_g=1\}}\,\, \int_{M }\,\mathrm{R}^{Per}(g)\,d\omega
$$
where $W^{1,2}(M)$ denotes the Sobolev space of functions on $(M,g)$ which, together with their first derivatives are square summable with respect to the Riemannian measure $d\mu_g$. If we set $\psi\,:=\,e^{\,-f/2}$, then one can characterize the functional  $\mathrm{F}[g]$ as  
the lowest eigenvalue $\lambda_1[g]$ of the Schr\"odinger-like operator on $(M, g)$ defined by
$$ -4\,\Delta_g\,\psi\,+\,\mathrm{R}(g)\,\psi\,=\,\lambda_1[g]\,\psi\;.$$
\\
The reader familiar with R. Hamilton's Ricci flow theory \cite{Ham1}, \cite{Ham2}, \cite{Ham3} will have certainly recognized in the above definitions some of the characters featuring in Perelman's celebrated proof \cite{18}, \cite{Per2}, \cite{Per3} of Thurston's geometrization program \cite{Thur1},\cite{Thur2}, \cite{Thur3}. This is  a direct consequence of the scaling and $\mathrm{Diff}(M)$--equivariance properties of the Ricci curvature which imply that not only Einstein, but also quasi--Einstein metrics do matter in Riemannian geometry.  Quasi--Einstein metrics are characterized by a Ricci tensor which can be written as
$$
\mathcal{R}ic(g)\,=\,\rho_{(g)}\,g\,-\frac{1}{2}\,\mathcal{L}_{V_{(g)}}\,g\,=\,\rho_{(g)}\,g\,-\frac{1}{2}\,\left(\nabla _i V_k\,+\,\nabla_k V_i\right)\;,
$$
for some constant $\rho_{(g)}$ and some complete vector field $V_{(g)}\in C^{\infty}(M,TM)$. If $V$ is a gradient, $V^i=g^{ik}\partial_k f$ for some $f\in C^\infty(M, \mathbb{R})$, then the quasi --Einstein condition becomes
$Ric^{B-E}(g,d\omega):=\,\mathcal{R}ic(g)\,+\,Hess_g\,f\,=\,\rho_{(g)}\,g$, \emph{i.e.} the isotropy of the Bakry--Emery Ricci curvature of the Riemannian manifold with density $(M,g, d\omega:=e^{-f}d\mu_g)$. More flavor to this remark is added if we note that quasi--Einstein metrics have a significant dynamical characterization. To this end, let us introduce a parameter $\beta$ with $0\leq\beta<\epsilon< \frac{1}{2\rho_{(g)}}$ and let us define $\lambda(\beta)\,:=\,(1-2\rho_{(g)}\,\beta)$. Consider \cite{derdzinski}  the one--parameter family of diffeomorphisms 
$\phi_\beta\,:M\longrightarrow  M$ solution of the non--autonomous ordinary differential equation 
$$\frac{\partial }{\partial \beta}\,\phi_\beta(p)\,=\, \frac{1}{
\lambda(\beta)}\,V_{(g)}(\phi_\beta(p)),\;\;\;\phi_{\beta=0}\,=\,id_M\;.$$
If we define the  one--parameter family of metrics $g(\beta)\,:=\,\lambda(\beta)\,\phi_\beta^*g$ \,($g(\beta=0)\,=\,g$), obtained by pulling back the metric $g$ under the action of the family of diffeomorphisms $\phi_\beta$ and rescalings $\lambda(\beta)$, then we can easily verify that, for $0\leq\beta<\epsilon< \frac{1}{2\rho_{(g)}}$,  the flow $\beta\longmapsto g(\beta)$ satisfies the evolution 
$$
\frac{\partial }{\partial \beta}\,g(\beta)\,=\,-2\,\rho_{(g(\beta))}\,g(\beta)\,+\,\mathcal{L}_{V_{(g(\beta))}}\,g(\beta)\,=\,-2\,\mathrm{R}ic(g(\beta))\;,
$$
with the initial condition $g(\beta=0)\,=\,g$. In other words, under the combined action of this family of diffeomorphisms and  scalings, the quasi--Einstein metric $g$ generates a self--similar solution  $g(\beta)\,:=\,\lambda(\beta)\,\phi_\beta^*\,g$,\, $0\leq\beta<\epsilon$,\, 
of the Ricci flow \cite{Ham1}
\begin{equation} 
\begin{tabular}{l}
$\frac{\partial }{\partial \beta }g_{ab}(\beta )=-2\,\mathrm{R}_{ab}(\beta ),$ \\ 
\nonumber\\ 
$g_{ab}(\beta =0)=g_{ab}$\, ,\,\,\,\,\, $0\leq \beta <\frac{1}{2\rho_{(g)}}$\,,%
\end{tabular}
\end{equation}
\textit{viz.}, quasi-Einstein metrics feature as those solutions of the Ricci flow which evolve only under the action of diffeomorphisms and scalings: the Ricci solitons \cite{hamiltonsolit}. 
\section{Ricci curvature, Wasserstein distance and the heat kernel}

The connection between Ricci curvature, diffeomorphisms, scalings, and Ricci flow comes further to the fore if we remove the requirement of absolute continuity of the measure $d\omega$ with respect to the reference Riemannian measure $d\mu_g$. We refer to any such manifold $(M, g, d\omega)$ as a \textit{weighted Riemannian manifold} and introduce the corresponding $\infty$--dimensional space $\mathcal{M}et(M)\times\,[Prob(M), d^W_g]$ of all weighted
  Riemannian manifolds , where $\mathcal{M}et(M)$ is the space of all smooth Riemannian metrics over $M$, and $Prob(M)$ denotes
the  space of all  Borel  probability measures $d\omega$ over $M$ endowed with the quadratic Wasserstein (or more appropriately, Kantorovich-Rubinstein) distance $d^W_g(d\omega_1, d\omega_2)$. Since this notion of distance plays a basic role in discussing Ricci curvature it is worthwhile to review its definition in some detail. Let ${\rm Prob}(M\times M)$ denote the set of Borel probability measures on the product space $M\times M$, and let us consider  the set of measures $d\sigma\in {\rm Prob}(M\times M)$  which reduce to $d\omega_1$ when restricted to the first factor and to $d\omega_2$  when restricted to the second factor, \emph{i.e.}
\begin{eqnarray}
\label{probprod}
&&{\rm Prob}_{\,\omega_1,\,\omega_2}\,(M\times M)\\
\nonumber\\
&&\,:=\, \left\{d\sigma \,\in {\rm Prob}(M\times M)\,
\left.  \right|\,\pi^{(1)}_\sharp \,d\sigma=d\omega_1,\, \pi^{(2)}_\sharp \,d\sigma=d\omega_2 \right\}\;,\nonumber
\end{eqnarray}
where $\pi^{(1)}_\sharp $ and $\pi^{(2)}_\sharp $ refer to the push--forward of  $d\sigma$ under  the projection maps $\pi^{(i)}$ onto the factors of $M\times M$. Measures $d\sigma \in {\rm Prob}_{\,\omega_1,\,\omega_2}\,(M\times M)$ are often referred to as \emph{couplings} between $d\omega_1$ and $d\omega_2$.   Given  a (measurable and non--negative) cost function $c\,:\,M\times M\rightarrow \mathbb{R}$, an optimal transport plan \cite{kantoro} $d\sigma_{opt}\in {\rm Prob}_{\,\omega_1,\,\omega_2}(M\times M)$ between the probability measures $d\omega_1$ and $d\omega_2$ \,$\in {\rm Prob}(M)$ is defined by the infimum, over all  $d\sigma(x, y)\in {\rm Prob}_{\,\omega_1,\,\omega_2}(M\times M)$, of the total cost functional
\begin{equation}
\int_{M\times M}\,c(x, y )\,d\sigma(x, y)\;.
\end{equation}
 On a Riemannian manifold $(M,g)$, the usual cost function is provided  \cite{lott1}, \cite{lott2}, 
\cite{mccan} by the squared Riemannian distance function $d^{\,2}_{\,g}(\cdot ,\cdot )$, and a major result of the theory \cite{mccan}, \cite{sturm}, \cite{brenier}, \cite{cordero},  is that  for any pair $d\omega_1$ and $d\omega_2$ $\in{\rm Prob}(M)$, there is an optimal transport plan $d\sigma_{opt}$, induced by a map $\Upsilon _{opt}:M\rightarrow M$ coming from a gradient.  The resulting expression for the total cost of the plan
\begin{equation}
\label{wassdist}
d_{\,g}^{\,W}\,(d\omega_1, d\omega_2)\,:=\,\left(\,\inf_{d\sigma\in {\rm Prob}_{\,\omega_1,\,\omega_2}(M\times M)} 
\;\int_{M\times M}\,d^{\,2}_{\,g}(x, y )\,d\sigma(x, y) \right)^{1/2}\;,
\end{equation}
characterizes the quadratic Wasserstein  distance between the two probability measures $d\omega_1$ and $d\omega_2$.
 Note that there can be distinct optimal plans $d\sigma_{opt}$ connecting general probability measures $d\omega_1$ and $d\omega_2$ $\in \rm{Prob}(M)$, whereas on the subset of $d\mu_g$--absolutely continuous measures the optimal transport plan is unique. The quadratic Wasserstein distance  $d_{\,g}^{\,W}$ defines a finite metric on ${\rm Prob}(M)$ and it can be shown that $({\rm Prob}(M),\,d_{\,g}^{\,W})$ is a geodesic space, endowed with the weak--* topology  (we refer to \cite{Villani}, \cite{savare}, \cite{VillaniON} for  the relevant properties of Wasserstein geometry and optimal transport we freely use in the following). Notice that if we denote  by $({\rm Prob}_{ac}(M),\,d_{\,g}^{\,W})$ the set of all $d\mu_g$--absolutely continuous Borel measures on the Riemannian manifold $(M,g)$, then the space of Riemannian manifolds with density, $\mathcal{M}et(M)\times\,[Prob_{ac}(M), d^W_g]$, is a dense subset of $\mathcal{M}et(M)\times\,[Prob(M), d^W_g]$.\\
\\
A deep rationale on Ricci curvature, scaling, $Diff(M)$-equivariance, and the induced Ricci flow  follows by observing that on the (dense) subset of Riemannian manifold with density,  the weighted Laplacian $\Delta^{(\omega)}_g\,:=\,\left(\Delta_g-\nabla{f}\cdot\,\nabla \right)$ is a symmetric operator with respect to the defining measure $d\omega$ and can be extended to a self--adjoint operator  in the space of square $d\omega$--summable functions $L^2(M, d\omega)$ generating the heat semigroup $e^{t\,\Delta^{(\omega)}}$,\, $t\,\in\,\mathbb{R}_{>0}$. The associated heat kernel $p^{(\omega)}_t(\cdot \, ,z)$ is defined as the minimal positive solution of
\begin{eqnarray}
\label{heatfloweight}
&&\left(\frac{\partial }{\partial  t}\,-\,\bigtriangleup _{\omega}\right)\,p^{(\omega)} _t(y, z)=0\;,\\
\nonumber\\
&&\lim_{t\searrow 0^+}\;p^{(\omega)} _t(y, z)\,d\omega(z)=\, \delta_{z}\;,\nonumber
\end{eqnarray}  
with $\delta_{z}$ the Dirac measure at $z\in (M,\,d\omega)$.\; The heat kernel $p^{(\omega)} _t(y, z)$ is $C^\infty$ 
on $\mathbb{R}_{>0}\times M\times M$, is symmetric $p^{(\omega)} _t(y, z)\,=\,p^{(\omega)} _t(z, y)$, satisfies the semigroup 
identity  $p^{(\omega)} _{t+s}(y, z)=\int_M\,p^{(\omega)} _t(y, x)p^{(\omega)} _s(x, z)\,d\omega(x)$, and $\int_M\,p^{(\omega)} _t(y, z)\,d\omega(z)=1$. Moreover, Varadhan's large deviation formula \cite{varadhan} holds 
\begin{equation}
\label{largedevweigh}
-\,\lim_{t\searrow 0^+}\,{t}\,\ln\,\left[\,p^{(\omega)}_{t}(y, z)\right]\,=\,
\frac{d_g^2(y, z)}{4}\;,
\end{equation}
where $d_g(y, z)$ is the Riemannian distance between the points $y$ and $z$ on $(M, g)$, and  the convergence is uniform over all $(M,g,\,d\omega)$. Since the map $(M, g)\longrightarrow (\mathrm{Prob}(M), d^W_{\,g})$ defined by $z\,\longmapsto\, \delta_z$ is an isometry, (one directly computes $d^W_{\,g}(\delta_y,\, \delta_z)\,=\,d_g(y,z)$, by using the trivial optimal plan $d\sigma(u,v)\,=\,\delta_y(u)\otimes \delta_z(v)$ in  (\ref{wassdist})), Varadhan's formula suggests that we can   use  the (weighted) heat kernel of $(M, g, d\omega)$, $(t,\delta _{x})\longmapsto p_t^{\omega}(\circ , x)$, with source at $x\in M$, to generate an injective embedding of $(M,g, d\omega)$ 
\begin{eqnarray}
\iota _{p_t^{\omega}}\,:\,(M,g)\,&\hookrightarrow &\,\left(\rm{Prob}\,(M),\,d_g^W\right) \nonumber\\
x\,&\longmapsto &\,\iota _{p_t^{\omega}}(x)\,:=\,p_t^{\omega}(\circ ,x)\,.\nonumber
\end{eqnarray} 
 in the space  $\left({\rm Prob}(M),\; d_{g, 2}^{\,W}\, \right)$  of
 all probability measures over $M$ endowed with  the quadratic Wasserstein distance, (see \cite{Carlo} for the pure Riemannian case and \cite{Mauro1} for the general case of the embedding of $(M,g, d\omega)$). If we exploit  this immersion, by pulling--back via $\iota _{p_t^{\omega}}$ the Wasserstein distance  $d_g^W\left(p_t^{\omega}(\circ ,x),\,\,p_t^{\omega}(\circ ,y) \right)$ to $M$, then   we get the $t$--dependent 
metric tensor on $M$ defined by 
$$
g_t(v(x), w(x)):=\,\int_{M_y}\,g^{ik}(y)\,\nabla _{i}^{(y)}\,\widehat{\psi}_{(t,x,v)}(y)\,\nabla _{k}^{(y)}\,\widehat{\psi}_{(t,x,w)}(y)\,p_t^{\omega}(y,x)\,d\mu_g(y)\;.
$$ 
where $\widehat{\psi}_{(t,v)}$,\ $\widehat{\psi}_{(t,W)}$ are the \emph{tangent vectors} in $\mathrm{Prob}(M)$ associated with the manifold tangent vectors $v$ and $w$. This procedure generates a scale dependent metric on 
$M$ such that $\lim_{t\searrow 0}\,g_t(v,v)\,=\,g(v,v)$,\, $v\in T_y\,M$, $y\in M$, and
$$
g_t(v,v)\,\leq\,e^{-\,2\,K_g^{B-E}\,t}g(v,v)\;,
$$
where $K_g^{B-E}$ denotes the lower bound of the Bakry--Emery Ricci curvature of $(M,g, d\omega))$. A rather sophisticated use of the (weak) Riemannian geometry of the Wasserstein space $\left(Prob(M,g),\,d_g^W) \right)$, (related to the Riemannian geometry of the diffeomorphisms group $\mathcal{D}iff(M)$ \emph{a' la Arnold} \cite{Arnold}), allows to compute \cite{Mauro1} a full--fledged flows for the metric $t\longmapsto g(t)$ and for the measure field $t\longmapsto f(t)$, \, \textit{viz.} 
$$\frac{\partial}{\partial t}\,f\,=\,-\,\bigtriangleup^{(z)} _\omega\,f$$
\begin{eqnarray}
\nonumber
&&\frac{\partial }{\partial t}\,g_t(u, w  )=\,-\,2\,{R}ic^{(t)}(u, w)
-\,2\,Hess\,f\,(u,v)\nonumber\\
\nonumber\\
&&\,-\,2\,\int_M\,\left(Hess\,\widehat{\psi}_{(t,u)}\cdot\,Hess\,\widehat{\psi}_{(t,w)}\right)\,
 p^{(\omega)} _t(y, z)\,d\omega(y)\nonumber\;,    
\end{eqnarray}
where ${R}ic^{(t)}$ denotes the Ricci curvature of the evolving metric $(M,\,g_t)$, and where $\widehat{\psi}_{(t,u)}$,\ $\widehat{\psi}_{(t,W)}$ are the \emph{tangent vectors} in $\mathrm{Prob}(M)$ representing the manifold tangent vectors $u$ and $w$, respectively. Hence we get an extended Ricci flow coupled 
with a (backward) parabolic evolution for  the measure $d\omega\,=\,e^{\,-f}\,d\mu_g$.

\section{Ricci curvature and quantum Riemannian geometry}

The connection between Ricci curvature, weighted heat kernel, Ricci flow and the  Wasserstein geometry of   $\mathcal{M}et(M)\times\,[Prob(M), d^W_g]$ can be extended to the discussion of the scale--dependent  random fluctuations of maps between  Riemann surfaces $(\Sigma, h)$ and weighted Riemannian manifolds $(M, g, d\omega)$.  This analysis can be used to describe in a mathematically rigorous way a quantum field theory (QFT) renormalization group perspective on Ricci curvature. 
With this remark we are coming to full circle.  Indeed, it is worth recalling that Quasi-Einstein metrics originated from theoretical physics 
\cite{DanThesis}, \cite{DanPRL}, \cite{DanAnnPhys}, in  the  perturbative analysis of the Renormalization Group for (Dilatonic) Non-Linear Sigma Model (NLSM), governed by an action which is  the quantum field theory avatar of  a harmonic map functional. To wit,  to a localizable\footnote{A suitable form of localizability is required since maps in $\mathrm{W}^{1,2}(\Sigma, M)$ may fail to have continuous representative. This delicate issue is discussed at length in \cite{Mauro1}} map $\phi\,:\,(\Sigma, h)\,\longmapsto \, (M, g, d\omega=e^{\,-f}d\mu_g)$ varying in the space of  Sobolev maps 
$\mathrm{W}^{1,2}(\Sigma, M)$ we associate  the action 
$$
{S}[\phi;\,\,a,\,f,\,g]\,:=\,
\frac{1}{2a}\,\int_{\Sigma}d^{2}x\,
\sqrt{h}\,\,\left[h^{\mu \nu }\partial _{\mu
}\phi^{i}\partial _{\nu }\phi^{k}\,\,
\,{g_{ik}}   +\,
2{a}\,\mathcal{K}_{h}(x)\,f(\phi(x))\right]
$$
where $\mathcal{K}_h$  is the Gauss curvature of the surface $(\Sigma,h)$,\, $a^{\,-\,1}\,g$ \, is the metric coupling associated with the (energy) scale parameter $a\,\in\,\mathbb{R}_{>0}$, and where the scalar function $
f: M\longrightarrow \mathbb{R}$\, characterizes the dilaton coupling of the model.  Heuristically,  quantum (actually, random) 
  fluctuations of
 $\phi:\Sigma \longrightarrow    M$ around a classical configuration $\phi_{cm}$, (typically, the center of mass of a large collection ($\rightarrow\infty$) of  constant maps $\{\phi_{(i)}\}$,\,identically and indipendently distributed with respect to  a sampling functional measure $\mathrm{D}[\phi]$ on 
$\mathrm{W}^{1,2}(\Sigma, M)$), can modify the  Riemannian geometry of  $(M, g, d\omega)$. The strategy is to exploit a scale dependent ($\beta:=\,a\,t$,\,$t\geq0$) perturbative renormalization to get a renormalization group (RG) flow for the metric and the dilaton couplings, $a^{-1}g$ and $f$,  controlled by a large deviation mechanism w.r.t. the Gaussian fluctuations around the (classical) background $\phi_{cm}$, (\emph{i.e.} by the control of the exponential decline of large field fluctuations, around $\phi_{cm}$, as the energy  scale  $\beta$ varies).
This procedure (re)constructs perturbatively the geometry in a ball around $\phi_{cm}$ as a function of the parameter $\beta$ according to \cite{DanThesis}, \cite{DanPRL}, \cite{DanAnnPhys}
$$
\frac{\partial }{\partial  \beta}\,g_{ik}(\beta)=-2\,R_{ik}(\beta) -2\nabla_i\nabla_k {f} \,-\,\frac{a}{2}\,(R_{ilmn}R_{k}^{lmn})+\,\mathcal{O}(a^{2})
$$
$$
\frac{\partial }{\partial  \beta}\,{f}(\beta)=\Delta {f}(\beta)-|\nabla {f}(\beta)|^2 \,+ \mathcal{O}(a^{2}).
$$
As long as we are in a weak coupling regime, characterized by the condition $a\left|\mathcal{R}m(g(\beta))\right|^{1/2}<<1$,
  we have the connection with Ricci flow in the DeTurck version \cite{DeTurck} 
$$
\frac{\partial }{\partial \beta }g_{ab}(\beta )\,=\,-2{R}_{ab}(\beta )\,-\,2\nabla_{a}\nabla_{b}\, {f},\;\;g_{ab}(\beta =0)=g_{ab} 
$$
coupled to the $d\omega$--weighted  forward heat equation for the dilaton
$$
\frac{\partial }{\partial \beta}{f} (\beta)=\Delta^{(\omega)} _{g(\beta)} {f}\;.  
$$
All this can be made mathematically rigorous either from the point of view of geometric analysis by exploiting Wasserstein geometry along the lines described above \cite{Mauro1} or, from the point of view of quantum theory, by framing the interplay between NLSM and Ricci flow into the algebraic quantum field theory approach \cite{Mauro2}. Question of space do not allow us to provide a more detailed description. The interested reader may find full details in the quoted papers.  It must be stressed that a full-fledged analysis of the  (weakly)
parabolic PDEs  featuring  in discussing Ricci flow  and RG flow requires a change of pespective
in the role of the dilaton coupling $f$. We need to impose \emph{Perelman's coupling}:\, \emph{viz.}\,we need to conjugate the dilaton ${f}$ rescaling to the $\beta$--evolving Riemannian measure $d\mu_{g(\beta)}$, by replacing ${f}\longrightarrow  \widetilde{f}$, with 
$\frac{\partial }{\partial \beta }\,e^{-\widetilde{f}(\beta )}d\mu _{g(\beta )}\, =\,0$, which couples the Hamilton-DeTurck Ricci flow above with the time--reversed, $\eta\doteq \beta ^{\ast }-\beta$, parabolic flow
$$\frac{\partial }{\partial \eta} \widetilde{f}(\eta)=\Delta _{g(\eta)} \widetilde{f}(\eta)-R(\eta)\,\widetilde{f}(\eta)\;
$$
In this way one recovers the celebrated monotonicity of Perelman's $F[g]$--energy along the Rici flow and its gradient--like nature.  This change of perspective is not confined to  the mathematical analysis of the  one-loop contribution (Ricci flow)  to the perturbative renormalzation group for NLSM, but also to the nature of  higher order terms such as  $a\,(R_{ilmn}R_{k}^{lmn})$, as discussed in detail in \cite{Mauro3}

\end{document}